\PassOptionsToPackage{colorlinks=true,linkcolor=blue,citecolor=blue,urlcolor=red}{hyperref}
\documentclass[journal,preprint]{vgtc} 

\onlineid{1584}

\preprinttext{Preprint. This manuscript has not been peer reviewed.}

\usepackage{amsmath}
\usepackage{amssymb}

\vgtcinsertpkg

\title{Beyond Mean Frametime: Time-Series Signatures for XR Timing Analysis}
\shortauthortitle{Th\"ans and Latoschik: Time-Series Signatures for XR Timing Analysis}

\author{Marvin Th\"ans\,\orcidlink{0009-0007-7131-193X} and Marc Erich Latoschik\,\orcidlink{0000-0002-9340-9600}}
\affiliation{University of W\"urzburg}
\authorfooter{%
  \item
  Marvin Th\"ans is with University of W\"urzburg.
  E-mail: marvin.thaens@uni-wuerzburg.de.
  \item
  Marc Erich Latoschik is with University of W\"urzburg.
}

\abstract{
XR systems expose timing quantities, such as motion-to-photon latency, frametime, or component-level runtime timings, that can be observed repeatedly as temporally ordered timing traces. Conventional reporting with means, standard deviations, percentiles, or histograms is useful, but it discards temporal ordering. We propose a general structure-aware methodology for analyzing and reporting XR timing traces. Each trace is represented by a compact, interpretable time-series signature, and collections of signatures can be visualized and compared statistically. We evaluate the method using engine-level application frametime traces from a large-scale in-the-wild VR dataset and compare timing signatures across HMD-labelled groups. Across multiple sampling and content-control conditions, structure-aware signatures reveal substantially stronger systematic multivariate differences between HMD-labelled groups than distribution-only summaries. A within-trace temporal-order shuffle control reduces this separation, particularly under content matching, providing direct evidence that original temporal ordering contributes information to the timing signatures. The strongest individual feature contributions vary across sampling and content-control conditions, indicating that no single timing characteristic dominates across analysis settings. Although demonstrated on application frametime, the representation operates on timing traces and therefore provides a basis for future application to other XR timing quantities, including instrumented motion-to-photon measurements.
}


\teaser{
  \centering
   \includegraphics[width=\linewidth, alt={Spectrogram of timing traces}]{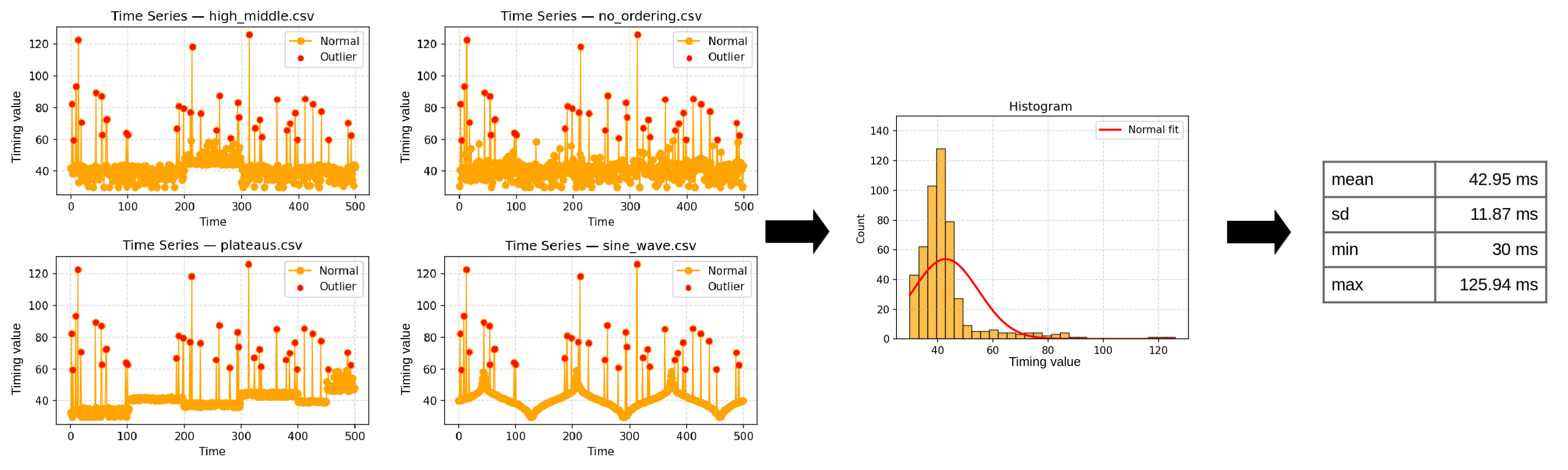}
     \caption{Different dynamics, same distribution, same reporting. The four traces (left) contain the same timing values within the time series, but arranged with different temporal structure. Reordering leaves the histogram unchanged (middle) and therefore yields identical distributional summaries (right), despite qualitatively different pacing patterns (bursts, dwell periods, oscillations).}
  \label{fig:teaser}
}

\graphicspath{{figs/}{figures/}{pictures/}{images/}{./}} 

\usepackage{tabu}                      
\usepackage{booktabs}                  
\usepackage{multirow}
\usepackage{enumitem}
\usepackage{orcidlink}
\usepackage{lipsum}                   
\usepackage{mwe}                       
\usepackage{pdfpages}

\usepackage{mathptmx}                  
\usepackage{placeins}

\usepackage{comment}
\usepackage[prologue,pdftex,dvipsnames,,svgnames]{xcolor}  
\usepackage{svg}
\usepackage{comment}
\usepackage{xargs}  
\usepackage[colorinlistoftodos,prependcaption,textsize=tiny]{todonotes}
\newcommandx{\unsure}[2][1=]{\todo[linecolor=red,backgroundcolor=red!25,bordercolor=red,#1]{#2}}
\newcommandx{\change}[2][1=]{\todo[linecolor=blue,backgroundcolor=blue!25,bordercolor=blue,#1]{#2}}
\newcommandx{\info}[2][1=]{\todo[linecolor=OliveGreen,backgroundcolor=OliveGreen!25,bordercolor=OliveGreen,#1]{#2}}
\newcommandx{\improvement}[2][1=]{\todo[linecolor=Plum,backgroundcolor=Plum!25,bordercolor=Plum,#1]{#2}}
\newcommandx{\thiswillnotshow}[2][1=]{\todo[disable,#1]{#2}}

\newcounter{extrefctr}
\newcommand{\extref}[1]{\refstepcounter{extrefctr}\href{#1}{\textcolor{red}{[\theextrefctr]}}}

\begin{document}

\firstsection{Introduction}
\maketitle
\label{sec:intro}

Timing is a core property of immersive XR systems because users interact with a closed sensorimotor loop. Several timing quantities characterize different parts of this loop. Motion-to-photon (MTP) latency describes the delay between user motion or input and the corresponding visual update on the display \cite{stauffertSimultaneousRunTimeMeasurement2020,warburtonMeasuringMotiontophotonLatency2023}, whereas application frametime characterizes application-side frame pacing at the engine/runtime boundary. Component-level timings may additionally describe individual stages such as tracking, rendering, compositing, or display update. These quantities are related, but they are not interchangeable: MTP latency is an end-to-end latency measure, while frametime describes application-level frame pacing.

Repeated observations of a timing quantity form a \emph{timing trace}, that is, a temporally ordered sequence of measurements. Such traces can be analyzed as \emph{time series}, allowing their temporal organization to be characterized using complementary time-series features. We refer to the resulting multivariate feature representation as a \emph{time-series signature} or \emph{timing signature}. This distinction separates the quantity being measured from the representation used to characterize and compare its temporal structure. Consequently, the same analysis approach can be applied to different XR timing quantities when suitable traces are available, while interpretation remains tied to the quantity that was actually measured.

Temporal ordering can encode variability, dependence, periodicity, transient structure, and burstiness. These properties are lost when a trace is reduced to a mean, standard deviation, percentile, or histogram. \Cref{fig:teaser} illustrates this limitation: the four traces contain identical values and therefore yield identical histograms and distributional summaries, yet differ markedly in how those values are organized over time. Distribution-only reporting cannot distinguish these pacing patterns.

This loss of temporal information is relevant because perceptual and interaction outcomes in XR are not determined only by average timing level. Added delay and timing instability can impair task performance \cite{teatherEffectsTrackingTechnology2009}, reduce presence \cite{meehanEffectLatencyPresence2003,slaterVirtualPresenceCounter2000}, and contribute to cybersickness through visual-vestibular mismatch \cite{davisSystematicReviewCybersickness2014,stauffertEffectsLatencyJitter2018,weechPresenceCybersicknessVirtual2019}. Controlled work further shows that jitter can impair interaction beyond mean delay \cite{stauffertEffectsLatencyJitter2018,teatherEffectsTrackingTechnology2009}, while sensitivity to motion and lag can depend on temporal frequency \cite{goldingMotionSicknessMaximum2001,kinsellaFrequencyNotAmplitude2016}. If perceptual and interaction effects depend on \emph{when} and \emph{how} variability unfolds \cite{goldingMotionSicknessMaximum2001,kinsellaFrequencyNotAmplitude2016}, then distribution-only reporting can obscure meaningful differences between devices, software stacks, or conditions. This is a practical problem for \emph{reliability} and \emph{reproducibility}: without characterizing temporal structure, it is difficult to compare systems fairly, to replicate timing-related findings across labs and platforms, or to determine whether apparently comparable conditions exhibit comparable timing.
More fundamentally, without structure-aware timing characterization, user outcomes such as performance, presence or cybersickness can be \emph{confounded}: observed effects may reflect the application design or content, but they may equally be driven (or amplified) by the underlying timing structure. These findings motivate characterizing not only the magnitude and distribution of timing values, but also their temporal organization.

Raw timing traces retain this temporal information but become difficult to inspect and compare across large collections. Scalar summaries are compact but discard temporal ordering. A compact structure-aware representation can bridge these two extremes by retaining information about complementary temporal characteristics while remaining suitable for comparison across systems, conditions, and studies.

Existing XR timing research provides established methods for measuring end-to-end and component-level timing \cite{dossenaOpenLDATSystemMeasurement2022,fristonMeasuringLatencyVirtual2014,heVideoBasedMeasurementSystem2000,stauffertSimultaneousRunTimeMeasurement2020,steedSimpleMethodEstimating2008}. The remaining gap addressed here lies downstream of measurement: once a timing trace has been obtained, common reporting practices largely reduce it to scalar or distributional features. What is missing is a compact, interpretable, and scalable representation that captures relevant aspects of temporal organization and supports systematic comparison across traces, conditions, and studies.

We address this representational gap by proposing a general methodology for structure-aware analysis and reporting of XR timing traces. The method treats each timing trace as a time series, extracts a compact set of interpretable time-series features, represents the trace as a multivariate timing signature, and then to visualize and compare the resulting signatures across traces, conditions, or systems. Our contribution is therefore representational rather than instrumental: we assume that a timing trace is available and provide a compact structure-aware layer for its analysis, visualization, and comparison.

We instantiate and validate the methodology using application frametime traces from the Open Extended Reality Recordings (BOXRR) dataset \cite{nairBerkeleyOpenExtended2024}, a large dataset of real-world VR Beat Saber recordings. In this empirical evaluation, the timing quantity is engine-level application frametime, traces are grouped by their recorded HMD label, and \texttt{catch22}/\texttt{catch24} features \cite{lubbaCatch22CAnonicalTimeseries2019} provide the concrete time-series signature implementation. Differences between HMD-labelled groups are used to evaluate whether timing signatures capture HMD-associated information that is less visible in conventional distribution-only summaries.

\textbf{Contributions:}
(i) \textbf{Structure-aware representation:} an XR timing analysis framework that represents timing traces with compact, interpretable time-series signatures that capture information about temporal structure discarded by distribution-only summaries \cite{fulcherHctsaComputationalFramework2017,lubbaCatch22CAnonicalTimeseries2019}.
(ii) \textbf{Reporting and visualization:} a scalable reporting layer for inspecting and comparing timing signatures across systems, conditions, and studies using standardized feature views, conventional absolute features where appropriate, and interpretable feature families \cite{aignerVisualizationTimeOrientedData2023}.
(iii) \textbf{Empirical validation:} an evaluation on more than five million engine-level frametime traces from real-world VR recordings, complemented by balanced sampling, content matching, and a within-trace temporal-order shuffle control that tests robustness and temporal-order sensitivity of the representation.

\section{Related Work}\label{sec:related}
Our work targets a representational gap between XR timing measurement and downstream comparison. Existing instrumentation can produce useful timing observations, but common reporting practice often collapses temporally ordered traces into distribution-only summaries. We therefore review XR timing measurement and reporting practices, then position structure-aware time-series features and visualizations as a compact reporting layer for comparing timing traces at scale.

\subsection{XR Timing Measurement and Reporting}
XR timing research commonly focuses on quantities such as MTP latency, jitter, and stage-specific runtime timings. End-to-end MTP latency is typically measured with dedicated instrumentation spanning tracking, simulation, rendering, compositing, and display update. Established approaches include video- and light-based rigs, synchronized sensing setups, and integrated measurement frameworks designed to improve temporal resolution and reproducibility \cite{dossenaOpenLDATSystemMeasurement2022,fristonMeasuringLatencyVirtual2014,heVideoBasedMeasurementSystem2000,stauffertSimultaneousRunTimeMeasurement2020,steedSimpleMethodEstimating2008}. These methods address how timing quantities are observed. Our work addresses how resulting timing traces can be represented and reported once observations are available.

For comparability, XR timing results are usually reported with scalar summaries such as mean, standard deviation, worst case, percentiles, histograms, or outlier counts. Such summaries are useful first-order features, and jitter-focused work has extended them with recursive modified-$z$-score outlier counts to highlight extreme deviations \cite{stauffertComparableEvaluationMethods2016}. However, prior XR work also argues that mean/worst-case optimization and distribution-only reporting can miss informative temporal patterns \cite{stauffertEffectsLatencyJitter2018,stauffertComparableEvaluationMethods2016}. Two timing traces can share the same distribution while differing in dependence, intermittency, oscillation, or transient structure (\Cref{fig:teaser}). This creates an XR-specific reporting gap: measurements may be temporally rich, but the reported features often discard temporal ordering.

\subsection{Structure-Aware Time-Series Analysis and Visualization}
Time-series analysis provides a general way to characterize temporal organization beyond distributional shape. Highly comparative time-series analysis (\textit{hctsa}) computes large collections of interpretable features and uses the resulting time-series$\times$feature matrix for systematic comparison across datasets and conditions \cite{fulcherHctsaComputationalFramework2017,fulcherHighlyComparativeTimeseries2013}. \textit{catch22} distills this idea into a compact canonical set selected for efficiency, low redundancy, and broad discriminative utility \cite{lubbaCatch22CAnonicalTimeseries2019}. Time-series data mining more broadly emphasizes that no single feature is universally best. Practical characterization typically combines features of distribution shape, dependence, periodicity, scaling, and predictability \cite{fuReviewTimeSeries2011}.

This perspective generalizes earlier pointwise outlier reporting. Whereas threshold-based counts describe how often samples deviate strongly from a typical level, structure-aware analysis can also describe dependence, recurrence, oscillatory organization, and local predictability. In anomaly-detection work, this distinction appears in methods that treat deviations from expected temporal structure as important, not only extreme values \cite{agrawalSurveyAnomalyDetection2015,blazquez-garciaReviewOutlierAnomaly2021,chandolaAnomalyDetectionSurvey2009,pangDeepLearningAnomaly2021}. Matrix Profile methods, for example, distinguish recurring subsequence motifs from rare discords \cite{yehMatrixProfileAll2016,yehTimeSeriesJoins2018}. The \texttt{catch22} feature set does not perform explicit motif/discord discovery, but related work such as C22MP shows that catch22-derived subsequence features can be combined with Matrix-Profile-style analysis \cite{tafazoliC22MPMarriageCatch222024}. For XR timing traces, these ideas motivate feature sets that capture more than sample amplitude.

Visualization is the second part of a scalable reporting layer. Time-oriented visualization work emphasizes that trace plots preserve ordering, distribution plots summarize variability, and aggregate matrix views support comparison when many traces must be inspected \cite{aignerVisualizationTimeOrientedData2023}. Reviews and interaction techniques further stress the need to balance scalability, pattern detection, and details-on-demand for large time-series collections \cite{hochheiserDynamicQueryTools2004,ortigossaTimeSeriesInformation2025}. Compact encodings such as horizon graphs illustrate how design choices affect the legibility of temporal variation \cite{SizingHorizonProceedings}.

Feature-based characterization connects these strands. Mapping each trace to a multivariate timing signature yields a compact trace$\times$feature matrix that can be inspected with heatmaps, clustered views, and feature-family groupings \cite{wilkinsonHistoryClusterHeat2009}. Such views retain interpretability because features can be discussed as descriptive families, such as distribution shape, extreme-event timing, autocorrelation, periodicity, symbolic dynamics, scaling, forecasting, or incremental differences \cite{lubbaCatch22CAnonicalTimeseries2019,fulcherHctsaComputationalFramework2017}. Multivariate tests such as MANOVA then provide a global comparison of whether signature vectors differ across groups while accounting for correlations among features \cite{andersonIntroductionMultivariateStatistical2003,obrienMANOVAMethodAnalyzing1985}. In this paper, we use these established tools as a structure-aware representation and reporting methodology for XR timing traces, not as a claim that any single feature family maps directly to a specific runtime mechanism.

\section{Dataset and evaluation conditions}\label{sec:dataset}
The empirical timing quantity in this evaluation is application/engine-level frametime derived from engine-internal timestamps. Frametime is not an MTP latency measurement: it reflects application-side frame pacing rather than the full tracking-to-display pipeline. We use frametime to instantiate and validate the proposed structure-aware methodology because it is available at large scale in the dataset. Each recording is associated with an HMD label identifying the reported headset/HMD system. We use this HMD label as the grouping variable for the empirical evaluation. The label does not characterize or control the complete XR configuration, including host hardware, operating system, drivers, or runtime configuration. Consequently, observed group differences are interpreted as HMD-associated differences rather than causal effects of the headset itself.

The Beat Saber traces analyzed in this work are derived from an extended version of the Berkeley Open Extended Reality Recordings (BOXRR) dataset \cite{nairBerkeleyOpenExtended2024}. Specifically, we use a newer release that extends the publicly available BOXRR-23 dataset and includes additional Beat Saber engine-level timestamp recordings not present in the publicly released BOXRR-23 dataset. The analysis presented here does not rely on BOXRR-specific annotations or metadata beyond the per-frame engine timestamps. We report results at an aggregate level and do not release raw traces. Subject to privacy and data-sharing constraints, the analysis pipeline is designed to transfer directly to the public BOXRR-23 release and comparable XR telemetry datasets.

We analyze per-frame engine timestamps from \textit{Beat Saber}, a timing-critical VR application in which performance depends on precise temporal coordination and motor control. The dataset comprises millions of in-the-wild gameplay recordings across multiple HMD-labelled groups and heterogeneous gameplay configurations, including songs, difficulty levels, modifiers, and environments. For each gameplay recording, we extract a frametime trace and compute time-series signatures per trace. Each trace captures observable frame pacing under real-world use, not component-resolved sub-latencies such as tracking, compositing, or display scan-out.

\subsection{HMD-labelled groups}
We compare the observable frametime characteristics associated with each recorded HMD label in-the-wild. The HMD label captures the reported headset category but does not control the complete system configuration. Our goal is therefore descriptive comparison of frametime characteristics across HMD-labelled groups rather than causal attribution to a single pipeline component.
Our analysis focuses on a core set of eight widely used head-mounted displays (HMD systems): Meta Quest~2, Quest~3, Quest~Pro, Rift~S, Valve Index, HTC Vive (original/dev/dvt), Vive~Pro, and Vive~Pro~2. We additionally include Meta Rift~CV1 and HTC Vive~Cosmos when they meet a minimum sample threshold of 50 traces, ensuring sufficient statistical support for HMD-group comparisons.

\subsection{Full dataset}
The raw dataset contains $N=7{,}949{,}374$ frametime traces spanning 32 distinct HMD labels. Because many labels are sparsely represented, we restrict our primary analysis to the 10 HMD groups with the highest sample counts. We also removed samples where the recorded playthrough covered less than 15\% of the song (i.e., incomplete trials). This yields the Full dataset condition used in our experiments, comprising $N=5{,}337{,}700$ traces. The Full condition reflects a naturalistic mix of songs, difficulty levels, environments, gameplay modes, and modifiers, and therefore captures the variability encountered in real-world use.
Filtering was applied in the following order: incomplete playthroughs were removed first, after which we retained the 10 HMD groups with the highest remaining sample counts.

As summarized in \Cref{tab:categorical-summary}, the dataset spans a wide range of content and gameplay conditions. Many of these factors (such as song, difficulty, environment, and gameplay modifiers) systematically influence frametime characteristics and can confound HMD-group comparisons if left uncontrolled.

\begin{table*}[t]
\centering
\fontsize{7}{7.5}\selectfont
\setlength{\tabcolsep}{1.1pt}
\renewcommand{\arraystretch}{1.0}
\def\tcell#1{\begin{minipage}[t]{\linewidth}\fontsize{7}{7.5}\selectfont\raggedright #1\end{minipage}}
\begin{tabular}{@{}
  >{\raggedright\arraybackslash}p{0.130\textwidth}
  >{\raggedright\arraybackslash}p{0.060\textwidth}
  >{\raggedright\arraybackslash}p{0.085\textwidth}
  >{\raggedright\arraybackslash}p{0.087\textwidth}
  >{\raggedright\arraybackslash}p{0.083\textwidth}
  >{\raggedright\arraybackslash}p{0.078\textwidth}
  >{\raggedright\arraybackslash}p{0.145\textwidth}
  >{\raggedright\arraybackslash}p{0.086\textwidth}
  >{\raggedright\arraybackslash}p{0.078\textwidth}
  >{\raggedright\arraybackslash}p{0.100\textwidth}@{}}
\toprule
\tcell{\textbf{HMD}\\\textbf{System}\\\textbf{(32)}} &
\tcell{\textbf{user}\\\textbf{(124,380)}} &
\tcell{\textbf{song}\\\textbf{name}\\\textbf{(65,489)}} &
\tcell{\textbf{difficulty}\\\textbf{(5)}} &
\tcell{\textbf{modes}\\\textbf{(25)}} &
\tcell{\textbf{modifier}\\\textbf{(643)}} &
\tcell{\textbf{speed}\\\textbf{(455)}} &
\tcell{\textbf{runtime}\\\textbf{(4)}} &
\tcell{\textbf{api}\\\textbf{(5)}} &
\tcell{\textbf{environment}\\\textbf{(42)}}\\
\midrule
\tcell{meta oculus\\quest 2\\(2,360,637)} &
\tcell{A\\(12,133)} &
\tcell{99.9\\(16,146)} &
\tcell{ExpertPlus\\(3,764,387)} &
\tcell{Standard\\(7,810,372)} &
\tcell{(empty)\\(6,657,014)} &
\tcell{0.0\\(7,945,842)} &
\tcell{steam\\(6,367,074)} &
\tcell{OpenVR\\(4,613,832)} &
\tcell{The First\\(3,289,853)}\\
\midrule
\tcell{meta oculus\\quest\\(gen 1)\\(2,011,252)} &
\tcell{B\\(9,621)} &
\tcell{Up \& Down\\(14,523)} &
\tcell{Expert\\(2,365,240)} &
\tcell{Lawless\\(95,024)} &
\tcell{NF\\(771,779)} &
\tcell{1.07e-38\\(744)} &
\tcell{oculus\\(1,335,835)} &
\tcell{Oculus\\(2,929,731)} &
\tcell{Big Mirror\\(1,226,430)}\\
\midrule
\tcell{valve index\\(1,612,857)} &
\tcell{C\\(9,176)} &
\tcell{Enemy\\(13,501)} &
\tcell{Hard\\(1,170,014)} &
\tcell{OneSaber\\(26,809)} &
\tcell{FS\\(78,805)} &
\tcell{1.65e-24\\(571)} &
\tcell{oculuspc\\(245,457)} &
\tcell{OpenXR\\(404,491)} &
\tcell{Nice\\(535,482)}\\
\midrule
\tcell{meta oculus\\rift s\\(551,995)} &
\tcell{D\\(8,095)} &
\tcell{666\\(12,827)} &
\tcell{Normal\\(417,161)} &
\tcell{360Degree\\(7,832)} &
\tcell{DA\\(62,306)} &
\tcell{1.83e-40\\(356)} &
\tcell{(empty)\\(1,008)} &
\tcell{Unknown\\(1,319)} &
\tcell{Panic\\(419,268)}\\
\midrule
\tcell{other\\(245,192)} &
\tcell{E\\(7,953)} &
\tcell{Idol\\(12,207)} &
\tcell{Easy\\(232,572)} &
\tcell{NoArrows\\(2,994)} &
\tcell{NA\\(57,960)} &
\tcell{1.352e-20\\(240)} &
\tcell{--} &
\tcell{(empty)\\(1)} &
\tcell{BTS\\(332,813)}\\
\bottomrule
\end{tabular}
\caption{Categorical variables in the raw dataset, with cardinality and most frequent levels. Many of these factors systematically affect frametime characteristics and can confound HMD-group comparisons if not controlled or matched.}
\label{tab:categorical-summary}
\end{table*}

\subsection{Restricted dataset (content-matched)}
To isolate HMD-associated differences from content effects, we additionally analyze a content-matched subset derived from the Full condition. This subset fixes song (\textit{Up \& Down}), difficulty (\textit{ExpertPlus}), environment (\textit{Nice}), gameplay mode (\textit{Standard}), modifier (none), and speed (0.0), yielding $N=6{,}740$ traces. By controlling these factors, directly addressing the confounding variables summarized in \Cref{tab:categorical-summary}, the Restricted condition enables a more direct comparison of HMD-associated frametime characteristics under identical application conditions and serves as a robustness check for the HMD-group differences observed in the Full condition.

\subsection{Balanced HMD-labelled subsets}
Because the dataset is highly imbalanced across HMD groups (\Cref{tab:categorical-summary}), we additionally perform analyses on balanced subsets in which each HMD group contributes an equal number of traces. For these runs, we randomly subsample each HMD group to match the smallest group size within the respective condition. In Full (bal.), this results in $N = 165{,}140$ total traces (10 HMD groups $\times$ 16{,}514 traces), while in Restricted (bal.), balancing yields a much smaller dataset with $N = 180$ total traces (10 HMD groups $\times$ 18 traces). In Restricted (bal.), we additionally enforce at most one trace per user so that each observation corresponds to a distinct user, eliminating within-user dependence in this condition.
Balanced subsets were generated with a fixed random seed, and the same sampled subsets were used for all feature-set comparisons.

This balancing removes differences in sample size as a potential driver of multivariate separation and provides a robustness check on HMD-associated timing-signature differences under both large-scale and small-sample conditions. Full (bal.) and Restricted (bal.) test whether MANOVA results are driven by unequal group sizes or large-$N$ effects.

\section{Methods}
\subsection{Timing-trace representation}
Let $x=(x_1,\dots,x_T)$ denote a timing trace: a temporally ordered sequence of observations of a timing quantity, represented as a time series. A timing signature is a feature vector $\phi(x)=(\phi_1(x),\dots,\phi_p(x))$ for one trace that summarizes selected properties of the time series, including distributional characteristics and order-dependent properties such as temporal dependence, periodicity, scaling, and frame-to-frame variability. All features are computed independently per trace.

In the present evaluation, $x$ contains per-frame application frametime values derived from engine-internal timestamps (\Cref{sec:dataset}). The broader method is not tied to frametime. Other timing quantities, such as instrumented MTP latency or component-level runtime timings, could be represented in the same way when suitable traces are available.

\subsection{Preprocessing and resampling}
The following preprocessing is specific to our frametime evaluation and the \texttt{catch22}/\texttt{catch24} implementation used here. The frametime traces are recorded at the application frame rate and therefore differ in sampling density across HMD groups. To reduce refresh-rate-induced differences in sampling density and because the \texttt{catch22} implementation expects evenly sampled input, we resample each per-frame series onto a common 1\,ms grid using zero-order hold (sample-and-hold). Concretely, each frame value is held constant until the next frame timestamp. This step places all traces on a common real-time grid, avoiding differences in the number of samples per unit time while preserving their piecewise-constant frametime representation. It does not eliminate cadence-related structure: features may still capture differences in frame-boundary density and pacing. Unlike interpolation-based resampling, it does not smooth between frames. Feature standardization for visualization is performed post hoc.
While this resampling places traces on a common sampling grid, time-series features may still reflect, time-series features may still reflect pacing- or cadence-related structure present in the observed traces. We retain this structure because delivered frame pacing is part of the observable timing represented by the trace.

\subsection{\texttt{catch22}-family feature extraction}
We instantiate the timing-signature representation with 22 canonical time-series characteristics from \texttt{catch22}, a compact and interpretable subset derived from the \texttt{hctsa} feature library \cite{fulcherHctsaComputationalFramework2017,lubbaCatch22CAnonicalTimeseries2019}. We compute these features with \texttt{pycatch22} version 0.4.5. The extracted features span multiple families defined in the source library, capturing distributional shape~(DN), linear and nonlinear temporal dependence via autocorrelation and mutual information (CO/IN), periodicity and spectral structure (PD/SP), symbolic dynamics~(SB), fluctuation and scaling properties~(FC/SC), and robust successive-difference measures~(MD). Feature family prefixes are encoded directly in the feature names (see \Cref{tab:catch24plus2-features}). \texttt{catch22}/\texttt{catch24} is the concrete implementation used in this paper, not the only possible implementation of structure-aware timing signatures.

\begin{table*}[t]
\centering
\small
\setlength{\tabcolsep}{4pt}
\renewcommand{\arraystretch}{1.05}
{\sloppy
\begin{tabular*}{\textwidth}{@{\extracolsep{\fill}}
  >{\raggedright\arraybackslash}p{0.04\textwidth}
  >{\raggedright\arraybackslash}p{0.15\textwidth}
  >{\raggedright\arraybackslash}p{0.30\textwidth}
  >{\raggedright\arraybackslash}p{0.17\textwidth}
  >{\raggedright\arraybackslash}p{0.19\textwidth}
  >{\raggedright\arraybackslash}p{0.04\textwidth}}
\toprule
\textbf{ID} & \textbf{Group} & \textbf{Feature name} & \textbf{Description} & \textbf{Used in feature set(s)} & \textbf{Source} \\
\midrule
DS\_1 & Distribution shape & \texttt{DN\_HistogramMode\_5}  & Histogram mode (5 bins)  & \texttt{catch22}\textbar\texttt{24}\textbar\texttt{24+2} & \cite{lubbaCatch22CAnonicalTimeseries2019} \\
DS\_2 & Distribution shape & \texttt{DN\_HistogramMode\_10} & Histogram mode (10 bins) & \texttt{catch22}\textbar\texttt{24}\textbar\texttt{24+2} & \cite{lubbaCatch22CAnonicalTimeseries2019} \\
EVT\_1 & Extreme event timing & \texttt{DN\_OutlierInclude\_p\_001\_mdrmd} & Positive outliers ($p=0.01$) & \texttt{catch22}\textbar\texttt{24}\textbar\texttt{24+2} & \cite{lubbaCatch22CAnonicalTimeseries2019} \\
EVT\_2 & Extreme event timing & \texttt{DN\_OutlierInclude\_n\_001\_mdrmd} & Negative outliers ($p=0.01$) & \texttt{catch22}\textbar\texttt{24}\textbar\texttt{24+2} & \cite{lubbaCatch22CAnonicalTimeseries2019} \\
LAC\_1 & Linear autocorrelation & \texttt{CO\_f1ecac} & ACF $1/e$ crossing time & \texttt{catch22}\textbar\texttt{24}\textbar\texttt{24+2} & \cite{lubbaCatch22CAnonicalTimeseries2019} \\
LAC\_2 & Linear autocorrelation & \texttt{CO\_FirstMin\_ac} & First minimum of ACF & \texttt{catch22}\textbar\texttt{24}\textbar\texttt{24+2} & \cite{lubbaCatch22CAnonicalTimeseries2019} \\
SS\_1 & Spectral summary & \texttt{SP\_Summaries\_welch\_rect\_area\_5\_1} & Welch power (band 5/1) & \texttt{catch22}\textbar\texttt{24}\textbar\texttt{24+2} & \cite{lubbaCatch22CAnonicalTimeseries2019} \\
SS\_2 & Spectral summary & \texttt{SP\_Summaries\_welch\_rect\_centroid} & Spectral centroid (Welch) & \texttt{catch22}\textbar\texttt{24}\textbar\texttt{24+2} & \cite{lubbaCatch22CAnonicalTimeseries2019} \\
PER\_1 & Periodicity & \texttt{PD\_PeriodicityWang\_th0\_01} & Periodicity score (Wang) & \texttt{catch22}\textbar\texttt{24}\textbar\texttt{24+2} & \cite{lubbaCatch22CAnonicalTimeseries2019} \\
SF\_1 & Simple forecasting & \texttt{FC\_LocalSimple\_mean1\_tauresrat} & Local forecast error ratio & \texttt{catch22}\textbar\texttt{24}\textbar\texttt{24+2} & \cite{lubbaCatch22CAnonicalTimeseries2019} \\
SF\_2 & Simple forecasting & \texttt{FC\_LocalSimple\_mean3\_stderr} & Local mean forecast error & \texttt{catch22}\textbar\texttt{24}\textbar\texttt{24+2} & \cite{lubbaCatch22CAnonicalTimeseries2019} \\
ID\_1 & Incremental differences & \texttt{MD\_hrv\_classic\_pnn40} & pNN40 successive differences & \texttt{catch22}\textbar\texttt{24}\textbar\texttt{24+2} & \cite{lubbaCatch22CAnonicalTimeseries2019} \\
SYM\_1 & Symbolic dynamics & \texttt{SB\_BinaryStats\_mean\_longstretch1} & Longest run above mean & \texttt{catch22}\textbar\texttt{24}\textbar\texttt{24+2} & \cite{lubbaCatch22CAnonicalTimeseries2019} \\
SYM\_2 & Symbolic dynamics & \texttt{SB\_BinaryStats\_diff\_longstretch0} & Longest run of decreases & \texttt{catch22}\textbar\texttt{24}\textbar\texttt{24+2} & \cite{lubbaCatch22CAnonicalTimeseries2019} \\
SYM\_3 & Symbolic dynamics & \texttt{SB\_MotifThree\_quantile\_hh} & 3-motif frequency (HH) & \texttt{catch22}\textbar\texttt{24}\textbar\texttt{24+2} & \cite{lubbaCatch22CAnonicalTimeseries2019} \\
SYM\_4 & Symbolic dynamics & \texttt{SB\_TransitionMatrix\_3ac\_sumdiagcov} & Transition self-covariance & \texttt{catch22}\textbar\texttt{24}\textbar\texttt{24+2} & \cite{lubbaCatch22CAnonicalTimeseries2019} \\
NLA\_1 & Nonlinear autocorrelation & \texttt{CO\_HistogramAMI\_even\_2\_5} & AMI from binned histogram & \texttt{catch22}\textbar\texttt{24}\textbar\texttt{24+2} & \cite{lubbaCatch22CAnonicalTimeseries2019} \\
NLA\_2 & Nonlinear autocorrelation & \texttt{CO\_trev\_1\_num} & Time-reversal asymmetry & \texttt{catch22}\textbar\texttt{24}\textbar\texttt{24+2} & \cite{lubbaCatch22CAnonicalTimeseries2019} \\
NLA\_3 & Nonlinear autocorrelation & \texttt{IN\_AutoMutualInfoStats\_40\_gaussian\_fmmi} & AMI (Gaussian, lag 40) & \texttt{catch22}\textbar\texttt{24}\textbar\texttt{24+2} & \cite{lubbaCatch22CAnonicalTimeseries2019} \\
OTH\_1 & Other / geometry & \texttt{CO\_Embed2\_Dist\_tau\_d\_expfit\_meandiff} & 2D embedding distance fit & \texttt{catch22}\textbar\texttt{24}\textbar\texttt{24+2} & \cite{lubbaCatch22CAnonicalTimeseries2019} \\
SAS\_1 & Self-affine scaling & \texttt{SC\_FluctAnal\_2\_rsrangefit\_50\_1\_logi\_prop\_r1} & Fluctuation analysis (R/S) & \texttt{catch22}\textbar\texttt{24}\textbar\texttt{24+2} & \cite{lubbaCatch22CAnonicalTimeseries2019} \\
SAS\_2 & Self-affine scaling & \texttt{SC\_FluctAnal\_2\_dfa\_50\_1\_2\_logi\_prop\_r1} & DFA scaling exponent & \texttt{catch22}\textbar\texttt{24}\textbar\texttt{24+2} & \cite{lubbaCatch22CAnonicalTimeseries2019} \\
RD\_1 & Raw distribution & \texttt{DN\_Mean} & Mean value & \texttt{catch24}\textbar\texttt{24+2}, \texttt{mean/SD only} & \cite{lubbaCatch22CAnonicalTimeseries2019} \\
RD\_2 & Raw distribution & \texttt{DN\_Spread\_Std} & Standard deviation & \texttt{catch24}\textbar\texttt{24+2}, \texttt{mean/SD only} & \cite{lubbaCatch22CAnonicalTimeseries2019} \\
RO\_1 & Raw outliers & \texttt{mad\_outlier} & MAD-based outlier score & \texttt{catch24+2}, \texttt{outlier only} & \cite{stauffertComparableEvaluationMethods2016} \\
RO\_2 & Raw outliers & \texttt{outlier\_level} & Outlier rate / level & \texttt{catch24+2}, \texttt{outlier only} & \cite{stauffertComparableEvaluationMethods2016} \\
\bottomrule
\end{tabular*}
}
\caption{Time-series features considered in this work, showing their feature family (group), identifiers, brief descriptions, membership in the derived feature sets, and source references.}
\label{tab:catch24plus2-features}
\end{table*}

In \texttt{catch22}, input time series are treated as evenly sampled sequences of values without explicit timestamps and are internally z-standardized for all features except mean and standard deviation. As a result, most \texttt{catch22} features emphasize relative distributional shape and temporal dependence rather than absolute scale or offset. Accordingly, absolute frametime level and dispersion in our feature sets are primarily carried by \texttt{DN\_Mean} and \texttt{DN\_Spread\_Std}, which are added in the \texttt{catch24} representation. The \texttt{catch24+2} set further augments these features with two MAD-based outlier measures (\texttt{mad\_outlier} and \texttt{outlier\_level}) following the recursive approach proposed by Stauffert et al.~\cite{stauffertComparableEvaluationMethods2016}. Because these outlier measures are scale-relative by construction, all features are used here in a descriptive, comparative sense to characterize frametime time-series structure rather than to define a single normative score or quality ranking.

To benchmark time-series characteristics against common reporting practice, we evaluate multiple feature sets: (i) \texttt{catch22} (22 time-series characteristics), (ii) \texttt{catch24}, which adds mean and standard deviation, (iii) \texttt{catch24+2}, which further adds two outlier measures used in comparable latency analyses \cite{stauffertComparableEvaluationMethods2016}, and two distribution-only summaries (mean/SD only, and outlier only).

\subsection{Feature standardization for visualization}
For visualization, each feature is standardized across the analyzed dataset using a z-score to place HMD-labelled groups on a common scale.

\subsection{MANOVA formulation}
Let $\mathbf{Y}\in\mathbb{R}^{N\times p}$ denote the matrix of time-series features for $N$ traces, with $p$ features per trace, and let $g$ denote the categorical HMD label. We use multivariate analysis of variance (MANOVA) to test the null hypothesis that the multivariate mean timing-signature vectors are identical across HMD-labelled groups. Statistical significance is assessed using Wilks' $\Lambda$ and Pillai's trace as global tests of group differences.
We report Pillai's trace as global measures of multivariate separation. The model is run independently for each feature set (\texttt{catch22}, \texttt{catch24}, \texttt{catch24+2}, mean/SD only, and outlier only) to contrast time-series characteristics and distribution-only summaries.

MANOVA is used here as a global test of whether multivariate frametime time-series characteristics differ across HMD-labelled groups, not as a tool for predicting outcomes or defining system rankings.
Because time-series feature distributions are not guaranteed to satisfy multivariate normality or homogeneity of covariance assumptions, we emphasize Pillai's trace, which is more robust under such violations, and complement parametric MANOVA results with permutation testing. This reduces sensitivity to assumption violations and unequal group sizes while preserving a global test of multivariate group differences.

In addition to global MANOVA, we perform pairwise multivariate comparisons between HMD groups using Hotelling's $T^2$ test with pooled covariance, converting $T^2$ to an $F$ statistic and applying Benjamini-Hochberg correction across all group pairs.

In addition, we conduct exploratory one-vs-rest univariate comparisons per feature using two-sample t-tests with effect sizes (Cohen's $d$ and $\eta^2$), reported only descriptively and corrected for multiple comparisons. Feature contributions are summarized by the median one-vs-rest $\eta^2$ across HMD groups within each analysis condition.

To complement parametric MANOVA test statistics, we additionally perform permutation tests using Pillai's trace as the test statistic. For each analysis condition and feature set, HMD labels are randomly permuted 7{,}000 times and the observed Pillai value is compared against the resulting null distribution. This provides a non-parametric robustness check that is less sensitive to violations of multivariate normality and unequal group sizes. Permutations are performed within each analyzed condition (Full, Full (bal.), Restricted, and Restricted (bal.)) using the same model specification.

\subsection{Robustness~analyses}
All MANOVA analyses are conducted across four dataset conditions: Full, Full (bal.), Restricted, and Restricted (bal.). Balanced conditions are created by randomly subsampling each HMD group to match the smallest group. These analyses assess robustness to both content variability and sample-size imbalance.

\subsection{Temporal-order shuffle control}
The baseline comparisons test whether time-series feature sets yield stronger HMD-group separation than mean/SD and outlier summaries, but they do not directly isolate the contribution of temporal ordering. We therefore perform a within-trace temporal-order shuffle control. For each selected trace, the raw per-frame frametime values are randomly permuted within the trace using a fixed reproducible seed. This preserves trace length and the set of raw observed frametime values while destroying their original sequential organization. The shuffled traces are then processed using the same 1\,ms ZOH preprocessing and \texttt{catch24}/\texttt{catch24+2} extraction pipeline as the original traces.

The control is evaluated on Full (bal.) and Restricted. The shuffled feature matrices are analyzed using the same MANOVA and pairwise Hotelling's $T^2$ procedures as the corresponding original conditions. This manipulation differs from the HMD-label permutation test above: label permutation tests the null hypothesis of no group association, whereas within-trace shuffling tests sensitivity of the extracted signatures to the original sequential organization of each trace.

\subsection{Feature heatmaps}
We visualize standardized features per HMD-labelled group as heatmaps to highlight which feature groups contribute most to HMD-associated differences. Features are grouped using the descriptive family labels in \Cref{tab:catch24plus2-features} (e.g., distribution shape (DS\_1-2), autocorrelation (LAC\_1-2, NLA\_1-3), spectral/periodic structure (SS\_1-2, PER\_1), symbolic dynamics (SYM\_1-4), scaling/forecasting (SAS\_1-2, SF\_1-2), and incremental differences (ID\_1)). This improves interpretability and supports mapping multivariate differences to specific time-series properties without implying a direct correspondence to underlying system mechanisms.

Analysis code, example data, and aggregated results supporting the statistical comparisons, tables, and figures are available at \url{https://go.uniwue.de/timing-signatures}.

\section{Results}
We evaluate the methodological validation question: does the proposed structure-aware representation capture systematic information that conventional distribution-only summaries fail to capture? Frametime traces grouped by recorded HMD label serve as the empirical validation case. We test whether time-series signatures yield stronger multivariate HMD-group separation than the distribution-only baselines, support interpretable visualization, and remain informative under content matching, balanced sampling, and within-trace temporal-order shuffling. We report global MANOVA separation, pairwise Hotelling's $T^2$ tests, and feature-level contributions.

\subsection{Structure-aware signatures vs.\ distribution baselines}
To evaluate the representational benefit of time-series signatures, we compare feature sets that capture aspects of temporal structure (\texttt{catch22}, \texttt{catch24}, \texttt{catch24+2}) against common distribution-only baselines (mean/SD only, outlier only). \Cref{tab:manova-hmd-main} summarizes Pillai's trace and p-values across analysis conditions, where higher Pillai's trace indicates stronger multivariate separation between HMD-labelled groups.

Across conditions, structure-aware signatures consistently yield stronger separation than distribution-only reporting. In the Full condition, mean/SD only provides almost no separation, whereas time-series feature sets retain substantial multivariate HMD-group separation. In the Restricted condition, mean/SD only becomes moderately informative once content variability is removed, but time-series signatures, especially those combining absolute and structural features (\texttt{catch24}, \texttt{catch24+2}), remain stronger in multivariate separation. Outlier only summaries remain weak and fail to distinguish HMD groups once content and sample size are controlled.

\subsection{Reporting/visualization: group-level timing signatures}
\begin{figure*}[t]
  \centering
  \includegraphics[width=\textwidth]{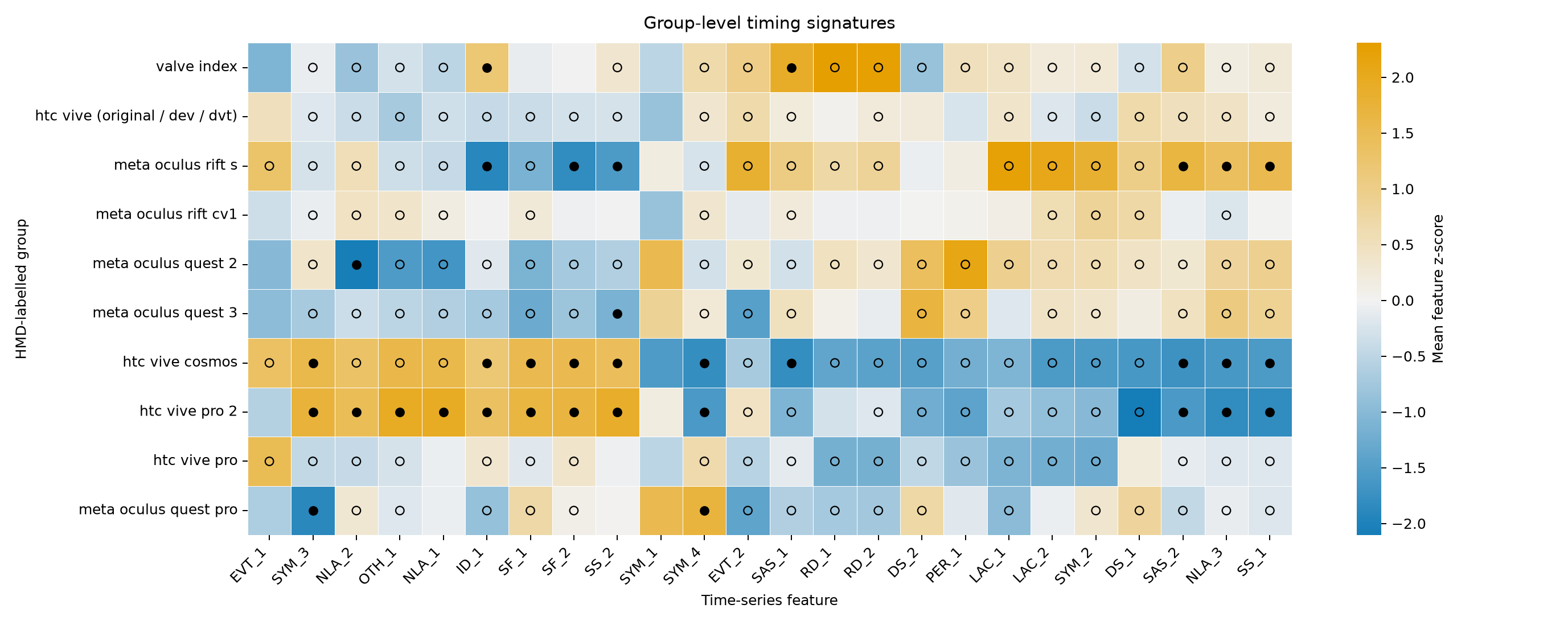}
  \caption{Dataset: Full (bal.) - Features: catch24.
Clustered HMD$\times$feature heatmap of per-HMD-group mean standardized feature values (group-level timing signatures; z-scores computed per feature across the dataset). Rows (HMD-labelled groups) and columns (features) are ordered by hierarchical clustering to group similar aggregate signatures. Circle markers are \emph{not} derived from clustering: they indicate per-cell \emph{one-vs-rest} tests for each (HMD group, feature) pair, using BH-FDR--adjusted $p$-values. Cells are marked significant if $p \le \alpha$ (default $\alpha=0.05$). Filled circles denote significant cells with $|d|\ge \delta$ (exceeding a practical-effect threshold; $\delta=0.25$), whereas open circles denote significant but small effects ($|d|<\delta$), where $d$ is the one-vs-rest standardized mean difference used for the contrast matrix.}
  \label{fig:hmd-feature-heatmap}
\end{figure*}

\Cref{fig:hmd-feature-heatmap} shows per-HMD-group mean standardized feature values as a z-normalized (per-feature) heatmap. Rows are HMD-labelled groups and columns are features. Each row is a group-level timing signature obtained by aggregating the trace-level timing signatures within an HMD-labelled group, whereas columns show which HMD groups are elevated or reduced on a specific timing property relative to the cohort mean. Coherent yellow/blue blocks therefore show which timing properties are elevated or reduced for an HMD group relative to the cohort mean, with yellow indicating above-average feature values and blue indicating below-average feature values. Circle markers denote one-vs-rest significance per cell, with filled vs.\ open circles indicating practically larger vs.\ smaller standardized differences. Read as a group-level signature view, the heatmap shows not only \emph{that} HMD groups differ, but also \emph{how}: whether a group is characterized more by frame-to-frame variability, temporal dependence, oscillatory structure, or level/dispersion. For example, HTC Vive Pro 2 and HTC Vive Cosmos show broad deviations across the same clustered feature block, whereas Meta Oculus Rift S shows a near-opposite pattern, indicating that groups differ in how frametime varies over time, not only in average level. The column view also highlights feature-specific contrasts: for instance, ID\_1 (pNN40 successive differences) is elevated for Valve Index and HTC Vive Pro 2 but reduced for Meta Oculus Rift S, indicating different amounts of frame-to-frame change across HMD groups.

\subsection{Validation: separation across HMD groups and analysis conditions}
\begin{figure}[t]
  \centering
  \includegraphics[width=\linewidth]{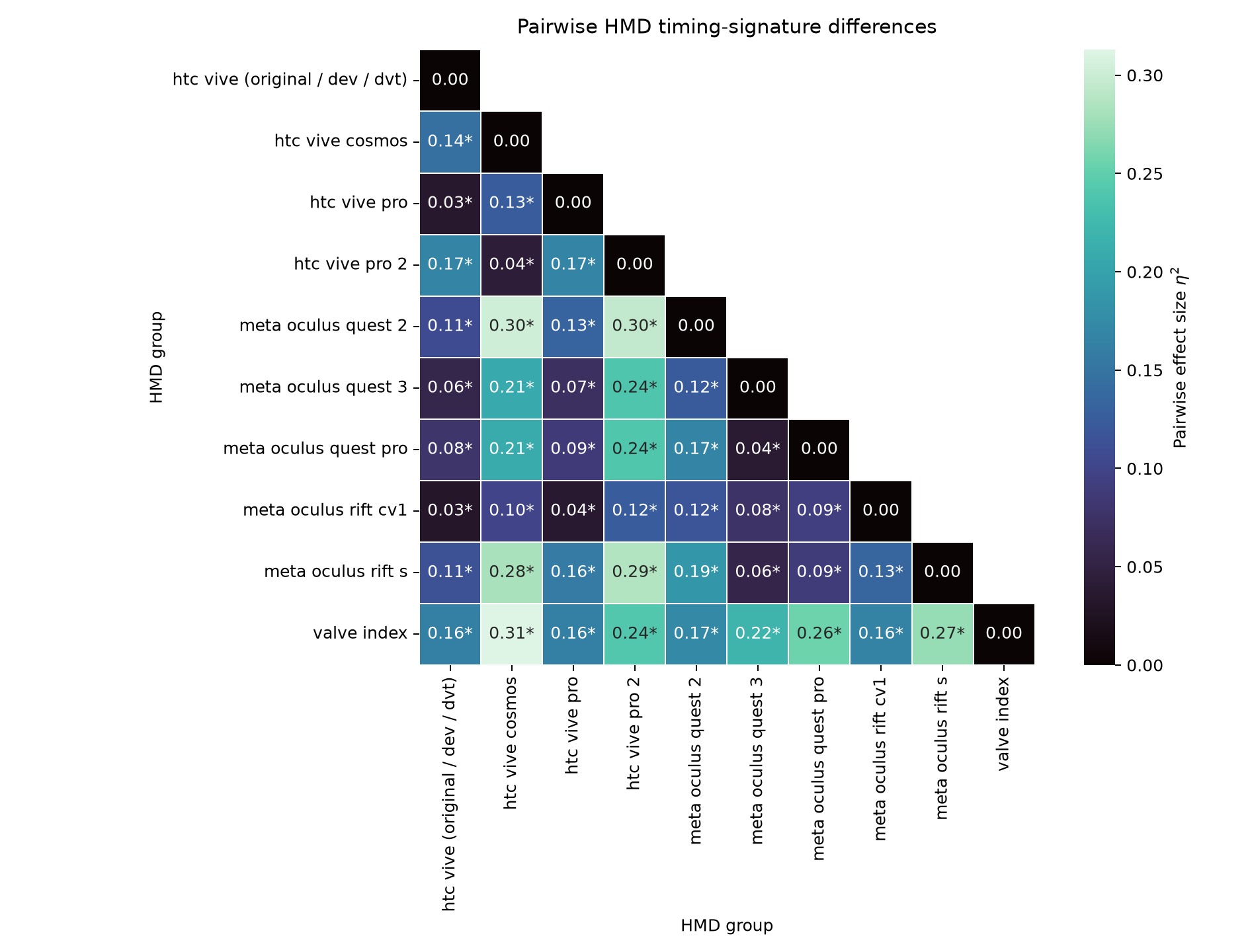}
  \caption{Pairwise question: Which HMD groups differ from which? Pairwise Hotelling's $T^2$ results for Full (bal.) using catch24 features. Hotelling's $T^2$ is the multivariate analogue of a two-sample $t$-test. Cell values show effect size ($\eta^2$); FDR-corrected significant comparisons ($p<0.05$) are marked with *.}
  \label{fig:hmd-pairwise-hotelling}
\end{figure}

\Cref{tab:manova-hmd-main} addresses the global question: \emph{is there any overall difference between HMD-labelled groups in multivariate timing-signature space?} Pillai's trace is the corresponding global separation score: larger values indicate stronger overall separation when all features are considered jointly. \Cref{fig:hmd-pairwise-hotelling} then addresses the pairwise question: \emph{which specific HMD groups differ from which?} Here Hotelling's $T^2$ is the multivariate analogue of a two-sample $t$-test, so each cell asks whether two groups $A$ and $B$ have the same multivariate mean signature, i.e., $H_0:\mu_A=\mu_B$.
We observe robust HMD-associated differences in multivariate timing-signature space, with the HMD-label factor remaining significant across Full, Full (bal.), Restricted, and Restricted (bal.).
\Cref{tab:manova-hmd-main} summarizes Pillai's trace and p-values for the primary feature sets across Full, Full (bal.), Restricted, and Restricted (bal.). 
Across analysis conditions, \texttt{catch24} and \texttt{catch24+2} provide the strongest overall separation, indicating that combining temporal structure with absolute distribution features (mean/SD) is most effective among the tested representations for HMD-group separation.
Relative to \texttt{catch22}, adding mean/SD (\texttt{catch24}) yields a larger practical gain than adding the two outlier features (\texttt{catch24+2}), which only marginally increases global separation/effect size and does not consistently increase the number of significant HMD-group pairs.
Pairwise counts in \Cref{tab:manova-hmd-main} are complementary to global MANOVA and are not expected to match it exactly. In Full and Full (bal.), pairwise counts largely saturate (often 45/45), showing that significance can be driven by very large sample size and therefore should be interpreted together with effect size. Median pairwise effect size (Med. $\eta^2$) clarifies practical magnitude: \texttt{catch24} and \texttt{catch24+2} are strongest overall, and \texttt{catch22} is consistently much larger than mean/SD only across all analysis conditions (e.g., Full: 0.044 vs 0.001, Full (bal.): 0.142 vs 0.005, Restricted: 0.129 vs 0.059, Restricted (bal.): 0.731 vs 0.274). These multivariate results are consistent with the signature heatmap (\Cref{fig:hmd-feature-heatmap}), where separation appears primarily row-driven: a subset of HMD-labelled groups exhibits distinct aggregate feature patterns across many (clustered) features, while others remain closer to the cohort mean. \Cref{tab:md_feature_robustness} then identifies which individual features contribute most strongly within each analysis condition.
Pairwise Hotelling's $T^2$ tests (\Cref{fig:hmd-pairwise-hotelling}) show extensive HMD-group pair separations for the catch24 feature set with strong multivariate separation, with large effect sizes persisting under balanced sampling, indicating robust HMD-associated signature differences rather than sample-size artifacts.

\begin{table}[t]
\centering
\scriptsize
\setlength{\tabcolsep}{2pt}
\begin{tabular*}{\columnwidth}{@{\extracolsep{\fill}}llcccc}
\toprule
\textbf{Dataset} & \textbf{Feature} & \textbf{Pillai} & \textbf{p} & \textbf{Pairs} & \textbf{Med. $\eta^2$} \\
\midrule
\multirow{5}{*}{\shortstack[l]{Full\\$N$=5,337,700}}
& \texttt{catch22}      & 0.237 & $<0.001$ & 45 & 0.044 \\
& \texttt{catch24}      & 0.259 & $<0.001$ & 45 & 0.045 \\
& \texttt{catch24+2}    & 0.271 & $<0.001$ & 45 & 0.049 \\
& mean/SD only          & 0.006 & $<0.001$ & 45 & 0.001 \\
& outlier only          & 0.016 & $<0.001$ & 45 & 0.002 \\
\midrule
\multirow{5}{*}{\shortstack[l]{Full (bal.)\\$N$=165,140}}
& \texttt{catch22}      & 0.292 & $<0.001$ & 45 & 0.142 \\
& \texttt{catch24}      & 0.313 & $<0.001$ & 45 & 0.145 \\
& \texttt{catch24+2}    & 0.337 & $<0.001$ & 45 & 0.162 \\
& mean/SD only          & 0.011 & $<0.001$ & 42 & 0.005 \\
& outlier only          & 0.022 & $<0.001$ & 44 & 0.007 \\
\midrule
\multirow{5}{*}{\shortstack[l]{Restricted\\$N$=6,740}}
& \texttt{catch22}      & 0.436 & $<0.001$ & 29 & 0.129 \\
& \texttt{catch24}      & 0.671 & $<0.001$ & 41 & 0.328 \\
& \texttt{catch24+2}    & 0.681 & $<0.001$ & 41 & 0.329 \\
& mean/SD only          & 0.258 & $<0.001$ & 35 & 0.059 \\
& outlier only          & 0.020 & $<0.001$ & 4 & 0.005 \\
\midrule
\multirow{5}{*}{\shortstack[l]{Restricted (bal.)\\$N$=180}}
& \texttt{catch22}      & 1.363 & $0.002$ & 6 & 0.731 \\
& \texttt{catch24}      & 1.904 & $<0.001$ & 25 & 0.870 \\
& \texttt{catch24+2}    & 2.031 & $<0.001$ & 27 & 0.903 \\
& mean/SD only          & 0.481 & $<0.001$ & 30 & 0.274 \\
& outlier only          & 0.131 & 0.173 & 0 & 0.064 \\
\bottomrule
\end{tabular*}
\caption{Global question: Is there an overall difference between HMD-labelled groups in multivariate timing-signature space? MANOVA summary across Full, Full (bal.), Restricted, and Restricted (bal.), with post-hoc pairwise comparison summary. Pillai's trace summarizes the strength of overall separation. \textit{Pairs} reports the number of significant different HMD-group pairs (Benjamini--Hochberg FDR, $q<0.05$, out of 45), and Med.~$\eta^2$ the median pairwise effect size from Hotelling's $T^2$ comparisons. Sample size $N$ is shown under each dataset label.}

\label{tab:manova-hmd-main}
\end{table}

Permutation tests based on Pillai's trace confirmed the parametric MANOVA results across analysis conditions and feature sets: all combinations were significant at $p<0.001$, except \texttt{catch22} in Restricted (bal.) ($p=0.0011$) and the \emph{outlier only} baseline in Restricted (bal.), which was not significant ($p=0.173$). This non-parametric check supports the robustness of the main multivariate separation pattern beyond MANOVA's parametric assumptions.

\subsection{Temporal-order shuffle control}
We next test whether destroying the original temporal ordering within each trace reduces the HMD-group information captured by the timing signatures. \Cref{tab:hmd-manova-shuffled-comparison} compares original and within-trace-shuffled frametime signatures for Full (bal.) and Restricted using the same MANOVA and pairwise Hotelling's $T^2$ procedures as above.

\begin{table*}[t]
\centering
\scriptsize
\setlength{\tabcolsep}{5pt}
\begin{tabular*}{\textwidth}{@{\extracolsep{\fill}}lllcccc}
\toprule
\textbf{Dataset} & \textbf{Ordering} & \textbf{Feature} & \textbf{Pillai} & \textbf{p} & \textbf{Pairs} & \textbf{Med. $\eta^2$} \\
\midrule
\multirow{4}{*}{\shortstack[l]{Full (bal.)\\$N$=165,140}}
& Original & \texttt{catch24}   & 0.313 & $<0.001$ & 45/45 & 0.145 \\
& Shuffled & \texttt{catch24}   & 0.247 & $<0.001$ & 45/45 & 0.130 \\
& Original & \texttt{catch24+2} & 0.337 & $<0.001$ & 45/45 & 0.162 \\
& Shuffled & \texttt{catch24+2} & 0.271 & $<0.001$ & 45/45 & 0.143 \\
\midrule
\multirow{4}{*}{\shortstack[l]{Restricted\\$N$=6,740}}
& Original & \texttt{catch24}   & 0.671 & $<0.001$ & 41/45 & 0.328 \\
& Shuffled & \texttt{catch24}   & 0.375 & $<0.001$ & 29/45 & 0.124 \\
& Original & \texttt{catch24+2} & 0.681 & $<0.001$ & 41/45 & 0.329 \\
& Shuffled & \texttt{catch24+2} & 0.386 & $<0.001$ & 29/45 & 0.134 \\
\bottomrule
\end{tabular*}
\caption{Temporal-order shuffle control. Comparison of original and within-trace-shuffled frametime signatures for Full (bal.) and Restricted. Shuffling destroys the original sequential organization of raw per-frame frametime values before the same 1\,ms ZOH preprocessing and feature extraction. Pillai's trace summarizes global HMD-group separation. Pairs gives FDR-significant Hotelling's $T^2$ comparisons out of 45, and Med.~$\eta^2$ gives the median pairwise effect size.}
\label{tab:hmd-manova-shuffled-comparison}
\end{table*}

In Full (bal.), within-trace shuffling reduces global and pairwise effect magnitude even though the number of significant pairs remains saturated at 45/45. For \texttt{catch24}, Pillai's trace decreases from 0.313 to 0.247 and median pairwise $\eta^2$ from 0.145 to 0.130. For \texttt{catch24+2}, Pillai's trace decreases from 0.337 to 0.271 and median pairwise $\eta^2$ from 0.162 to 0.143. Because Full (bal.) contains approximately 165k traces, the unchanged pair count should not be interpreted as absence of a shuffle effect. Reductions in Pillai's trace and median pairwise effect size are the more informative measures here.

The reduction is substantially larger under content matching. In Restricted, shuffling \texttt{catch24} decreases Pillai's trace from 0.671 to 0.375, significant pairwise HMD-group comparisons from 41/45 to 29/45, and median pairwise $\eta^2$ from 0.328 to 0.124. For \texttt{catch24+2}, Pillai's trace decreases from 0.681 to 0.386, significant pairs again decrease from 41/45 to 29/45, and median pairwise $\eta^2$ decreases from 0.329 to 0.134. When major content variables are controlled, destroying original within-trace ordering therefore removes a substantial portion of the multivariate HMD-group separation captured by the signatures. Separation remains statistically significant after shuffling, so temporal ordering is not the sole source of HMD-associated signature differences.

Together, these results provide direct evidence that the original sequential organization of frametime contributes to the HMD-associated information captured by the timing signatures. The remaining separation after shuffling shows that temporal ordering is one contributor rather than the sole source of group differences.

\subsection{Additional findings: feature contributions}
To contextualize which individual characteristics shift most across HMD groups, we rank features by the median one-vs-rest $\eta^2$ across HMD groups within each analysis condition. The strongest contributors vary by condition: \texttt{NLA\_2} leads in Full, \texttt{ID\_1} in Full (bal.), \texttt{SF\_1} in Restricted, and \texttt{RD\_1} in Restricted (bal.) (\Cref{tab:md_feature_robustness}).
The other high-ranking features span incremental-difference, scaling, forecasting, spectral, and symbolic families. Given the large sample sizes, we emphasize effect sizes and condition-specific rankings over $p$-values when interpreting individual features.

\begin{table}[t]
\centering
\scriptsize
\setlength{\tabcolsep}{5pt}
\begin{tabular*}{\columnwidth}{@{\extracolsep{\fill}}lllc}
\toprule
\textbf{Dataset} & \textbf{Rank} & \textbf{Feature} & $\boldsymbol{\eta^2}$ \\
\midrule
\multirow{3}{*}{Full}
& 1 & \texttt{NLA\_2} & 0.012 \\
& 2 & \texttt{ID\_1}  & 0.008 \\
& 3 & \texttt{SAS\_2} & 0.005 \\
\midrule
\multirow{3}{*}{Full (bal.)}
& 1 & \texttt{ID\_1}  & 0.011 \\
& 2 & \texttt{SF\_1}  & 0.006 \\
& 3 & \texttt{SS\_1}  & 0.005 \\
\midrule
\multirow{3}{*}{Restricted}
& 1 & \texttt{SF\_1}  & 0.037 \\
& 2 & \texttt{SS\_1}  & 0.020 \\
& 3 & \texttt{SS\_2}  & 0.017 \\
\midrule
\multirow{3}{*}{Restricted (bal.)}
& 1 & \texttt{RD\_1}  & 0.030 \\
& 2 & \texttt{SYM\_3} & 0.027 \\
& 3 & \texttt{SS\_2}  & 0.017 \\
\bottomrule
\end{tabular*}
\caption{Top-3 univariate feature contributors for the \texttt{catch24} feature set across analysis conditions, ranked by the median one-vs-rest effect size $\eta^2$ across HMD groups. The rankings vary across conditions, with \texttt{NLA\_2}, \texttt{ID\_1}, \texttt{SF\_1}, and \texttt{RD\_1} leading in Full, Full (bal.), Restricted, and Restricted (bal.), respectively.}
\label{tab:md_feature_robustness}
\end{table}

The condition-specific rankings indicate that HMD-associated frametime differences involve multiple complementary characteristics rather than a single dominant feature. Forecasting and spectral features are prominent under content matching, while raw mean becomes strongest only in Restricted (bal.).

\section{Discussion}
Our results support the central methodological premise of this work: XR timing traces can contain systematic temporal structure that is not captured by distribution-only summaries. In the empirical validation, application frametime traces from HMD-labelled groups remain separable in multivariate time-series signature space under content matching and balanced sampling, whereas conventional mean/SD and outlier-only summaries are substantially weaker in several conditions. This conclusion is supported not only by stronger separation of the structure-aware feature sets, but also by the within-trace temporal-order shuffle control: destroying original sequential organization reduces HMD-group separation in both tested cohorts, with a particularly strong reduction under content matching.

\subsection{What structure-aware timing signatures capture}
Interpreting trace-level timing signatures through feature families provides a descriptive vocabulary for \emph{which temporal properties} differ across traces and, after aggregation, across groups. These families describe the type of time-series property a feature quantifies.

In our feature sets, \emph{distribution shape} and \emph{raw distribution} features summarize frametime level, dispersion, and tail behavior. \emph{Extreme-event timing} features characterize the temporal occurrence of positive and negative extremes, whereas the separately included raw-outlier features summarize deviations using the recursive MAD-based procedure. \emph{Linear} and \emph{nonlinear dependence} features characterize short-term temporal dependence and characteristic time scales. \emph{Spectral summary} and \emph{periodicity} features capture oscillatory structure and \emph{self-affine scaling}, \emph{simple forecasting}, and \emph{incremental-difference} features describe longer-range fluctuation structure, local predictability, and frame-to-frame variability. No single feature leads across all analysis conditions (\Cref{tab:md_feature_robustness}): nonlinear asymmetry leads in Full, incremental differences in Full (bal.), local forecasting in Restricted, and raw mean in Restricted (bal.).

Although MANOVA is reported primarily as a global test, the separation appears structured rather than diffuse. The feature-wise rankings (\Cref{tab:md_feature_robustness}) and heatmap patterns (\Cref{fig:hmd-feature-heatmap}) show condition-dependent contributions from nonlinear dependence, incremental differences, forecasting, spectral, scaling, symbolic, and raw-distribution features. This variation cautions against treating any individual descriptor as universally dominant.

\subsection{Reporting and comparing XR timing traces}
Aggregating trace-level timing signatures into a standardized group-level heatmap provides a scalable reporting layer for timing structure. Raw traces remain important for inspecting individual bursts, oscillations, or stalls, but they do not scale to millions of recordings. Scalar summaries scale, but discard ordering. A group-level timing-signature heatmap occupies the middle ground: it gives readers a compact view of which cohorts share similar multivariate structure and which feature families drive the differences.

This supports a practical comparison workflow. A researcher or developer can collect timing traces for the systems or conditions under study, compute conventional summaries and time-series signatures, inspect whether groups separate in standardized feature space, identify the feature families responsible for separation, and report whether the observed difference is primarily distributional or temporal. The method is therefore comparative and diagnostic rather than a universal quality score.

\subsection{Validation through HMD-group differences}
HMD-labelled groups provide the empirical validation case for the representation. The persistence of group differences under balancing and content matching demonstrates that trace-level timing signatures contain HMD-associated information beyond conventional distributional summaries.

The shuffle control operationalizes the conceptual example in \Cref{fig:teaser}: it changes the ordering of observed raw frametime values while retaining the raw values themselves, allowing us to test whether sequential organization contributes to downstream signature differences. In Full (bal.), pairwise significance remains saturated at 45/45 because this large cohort has very high statistical power, but Pillai's trace and median pairwise effect size both decrease after shuffling. The pair count therefore has a ceiling effect and is less informative than the effect-size changes in this cohort.

The shuffle effect is much more pronounced in Restricted, where major content variables are matched. Global separation decreases strongly, median pairwise effect sizes fall strongly, and significant pairwise HMD-group comparisons drop from 41/45 to 29/45 for both \texttt{catch24} and \texttt{catch24+2}. This supports the interpretation that temporal ordering contributes particularly strongly to the observed signature differences once content-driven heterogeneity is reduced. Because shuffled signatures remain significantly different across HMD groups, group separation is not exclusively temporal. Absolute/distributional characteristics and other information captured by the preprocessing and signature representation also contribute.

The contrast between the Full and Restricted conditions is practically important. Real-world XR studies and benchmarks often differ in HMD systems, applications, content, and runtime setups. In the Full (bal.) analysis, mean/SD yields almost no multivariate separation, whereas time-series signatures reveal HMD-associated signature differences (\Cref{tab:manova-hmd-main}). In the Restricted (bal.) condition, the outlier-only baseline becomes non-significant and shows no significant pairwise separations, indicating that extreme-value summaries alone are insufficient once major confounds are controlled. Because very large sample sizes can produce widespread significance, we interpret $p$-values together with effect sizes and pairwise summaries.

\subsection{Practical implications for XR experiments and cross-study comparison}
The proposed signatures can be used as a pre-study screening and reporting tool. If two HMD groups have similar mean frametimes but clearly different timing signatures, the HMD choice should be treated as a potential timing confound rather than an interchangeable implementation detail. If separation is driven by dependence, periodicity, or incremental variability, authors can report those temporal properties alongside mean/SD instead of reducing the trace to distributional summaries.

The heatmap patterns illustrate this use. HTC Vive Pro 2 and HTC Vive Cosmos show deviations across a similar clustered feature block, suggesting related group-level timing signatures beyond average frametime level. Meta Oculus Rift S shows a near-opposite pattern in the same view. The practical implication is not that one HMD group is universally better, but that cross-system comparisons using these HMD labels should either control for HMD group, report the signature difference explicitly, or avoid attributing downstream behavioral outcomes solely to application content.

\subsection{Limitations and scope}
This study evaluates the methodology on engine-level application frametime traces, not on end-to-end MTP latency. Frametime reflects application-side frame pacing and does not directly capture tracking, compositing, or display pipeline delays. The HMD label captures the reported headset category but does not control the complete hardware and software environment. Observed differences should therefore be interpreted as HMD-associated timing differences, not causal effects of the headset itself.

Although content matching and balanced analyses mitigate major confounds, the dataset remains heterogeneous with respect to hardware configurations, runtime/driver stacks, and execution environments. In the Restricted (bal.) condition each user contributes at most one trace, but in the Full condition users may contribute multiple trials. Accordingly, the reported $p$-values should be interpreted primarily as evidence of systematic differences in this dataset rather than as population-level causal estimates.

Preprocessing choices can influence structure features. We resample frametime traces onto a common 1\,ms grid using zero-order hold to improve cross-system comparability in real time. This representation is piecewise constant between frame boundaries and can make certain short-lag and incremental-difference features sensitive to the density of frame updates (effective delivered cadence) as well as the magnitude of step changes. We treat cadence sensitivity as part of observable frame pacing, but sensitivity to grid resolution and resampling scheme has not been exhaustively quantified.

The shuffle control was evaluated only for Full (bal.) and Restricted, not for all four primary analysis conditions. It preserves the raw per-frame values and removes its original ordering before applying the same ZOH preprocessing. It should therefore be interpreted as an ordering removal of the complete analysis pipeline rather than as proof that every marginal property of the resampled signal is invariant.

Finally, MANOVA provides a global test of multivariate differences, and pairwise Hotelling's $T^2$ tests identify which group pairs differ in multivariate timing-signature space. These tests do not uniquely decompose correlated multivariate differences into per-feature causes. Family-level interpretations are descriptive and should not be read as causal attribution without additional instrumentation and controlled interventions. Mean/SD also remain easier to grade because prior work already links them to user outcomes, whereas an analogous reference frame for time-series features is still missing.

\subsection{Future validation across timing quantities}
A natural next step is to apply the same representational pipeline to independent datasets and additional timing quantities, especially instrumented MTP traces. Uniformly sampled MTP measurements would remove the need for frametime-specific ZOH resampling and help separate cadence effects from step-magnitude dynamics. Mixed-reality and augmented-reality workloads with heavier tracking and compositing demands also require validation before claiming generality beyond the VR frametime case studied here.

Perceptual validation is equally important. The current signatures describe selected temporal properties. They do not establish which individual dimensions matter to users. Future studies could test whether mean-matched traces with different timing signatures affect smoothness, comfort, task performance, presence, or cybersickness. Such work would turn the present reporting layer into a bridge between XR timing measurement and perceptual or behavioral outcomes.

\section{Conclusion}
We presented a structure-aware methodology for analyzing and reporting XR timing traces. The central idea is to separate the timing quantity being observed from the representation used for comparison: repeated observations form a timing trace, the trace is represented as a time series, and compact time-series features summarize it as a timing signature. This representation complements conventional means, standard deviations, percentiles, and histograms by capturing selected properties of temporal organization that distribution-only summaries discard.

We validated the approach on a large real-world dataset of engine-level VR frametime traces. In this evaluation, HMD-labelled groups provide the group structure, and \texttt{catch22}/\texttt{catch24} features provide the concrete signature implementation. The results show that structure-aware signatures reveal systematic multivariate differences that are much less visible in distribution-only summaries and remain observable under content matching and balanced sampling. A within-trace temporal-order shuffle control further reduced HMD-group separation, particularly under content matching, showing that original temporal organization contributes to these signature differences rather than the separation arising solely from unordered frametime values. The remaining post-shuffle separation indicates that temporal ordering is one contributor rather than the sole source of HMD-associated signature differences.

The method currently supports comparison and reporting rather than an absolute quality score. It does not identify causal hardware or runtime mechanisms, and this paper does not validate the representation on MTP latency, AR, or MR traces. Future work should apply the same pipeline to instrumented MTP measurements and connect individual timing-signature dimensions to perceptual and behavioral outcomes.

Overall, the contribution is not a new latency metric or an HMD classifier, but a structure-aware representation and reporting layer between XR timing measurement and subsequent system, experimental, or perceptual analysis.

\acknowledgments{
Generative AI was used solely for technical assistance with LaTeX template conversion and manuscript formatting. Specifically, the Codex CLI using the GPT-5.5 model (OpenAI, https://openai.com/) was used for these tasks. DeepL (DeepL SE, web service, https://www.deepl.com/) and Grammarly (Grammarly, Inc., web service, https://www.grammarly.com/) were used for language translation, rewriting, grammar correction, and stylistic polishing. These tools were not used to perform scientific analyses, generate or modify data or results, or determine the scientific interpretation or conclusions of the manuscript. All resulting changes were reviewed and verified by the authors.
}

\section*{Data Availability Statement}
The data analyzed in this study derive from an extended release of the Berkeley Open Extended Reality Recordings (BOXRR) dataset. At the time of submission, this extended release is being prepared as a separate data publication by the dataset maintainers and is not yet publicly available. The authors do not redistribute raw timing traces in this article. Access to the extended dataset will be provided through the forthcoming dataset release, subject to participant privacy, consent, and data-sharing constraints. The analysis code, example data, and aggregated results supporting the reported statistical comparisons, tables, and figures are publicly available at \url{https://go.uniwue.de/timing-signatures}. The repository does not redistribute the raw BOXRR timing traces.

\bibliographystyle{abbrv-doi-hyperref-narrow}

\bibliography{main}

\appendix 

\end{document}